\documentclass[floats,floatfix,showpacs,amssymb,prd,twocolumn,superscriptaddress,nofootinbib,nolongbibliography,reprint]{revtex4-1}

\usepackage{amssymb,amsmath,verbatim,mathtools,needspace,enumitem,etoolbox,graphicx,physics,microtype,afterpage,bigints,gensymb,tabularx}
\usepackage[dvipsnames, usenames]{xcolor}
\definecolor{linkcolor}{rgb}{0.0,0.3,0.5}
\definecolor{dodgerblue}{HTML}{1E90FF}
\usepackage[unicode, colorlinks=true, linkcolor=linkcolor, citecolor=linkcolor, filecolor=linkcolor,urlcolor=linkcolor, pdfusetitle]{hyperref}
\usepackage[all]{hypcap}
\usepackage[T1]{fontenc}
\usepackage[utf8]{inputenc}
\usepackage{orcidlink}
\renewcommand{\vec}[1]{\boldsymbol{#1}}
\newcommand{\beq}{\begin{eqnarray}}
\newcommand{\eeq}{\end{eqnarray}}
\newcommand{\del}{\partial}

\makeatletter
\newcommand*{\balancecolsandclearpage}{\close@column@grid \cleardoublepage \twocolumngrid}
\makeatother

\usepackage{lipsum}
\newcommand{\jhu}{\affiliation{William H. Miller III Department of Physics and Astronomy, Johns Hopkins University, Baltimore, Maryland 21218, USA}}

\begin{document}

\title{The impact of confusion noise on golden binary neutron-star events \\ in next-generation terrestrial observatories}

\author{Luca Reali$\,$\orcidlink{0000-0002-8143-6767}}
\jhu

\author{Andrea Antonelli$\,$\orcidlink{0000-0001-6536-0822}}
\jhu

\author{Roberto Cotesta$\,$\orcidlink{0000-0001-6568-6814}}
\jhu

\author{Ssohrab Borhanian$\,$\orcidlink{0000-0003-0161-6109}}
\affiliation{Theoretisch-Physikalisches Institut, Friedrich-Schiller-Universit{\"a}t Jena, 07743, Jena, Germany}

\author{Mesut \c{C}al{\i}\c{s}kan$\,$\orcidlink{0000-0002-4906-2670}}
\jhu

\author{Emanuele Berti$\,$\orcidlink{0000-0003-0751-5130}}
\jhu

\author{B. S. Sathyaprakash$\,$\orcidlink{0000-0003-3845-7586}}
\affiliation{Institute for Gravitation and the Cosmos, Department of Physics, Penn State University, University Park, Pennsylvania 16802, USA}
\affiliation{Department of Astronomy and Astrophysics, Penn State University, University Park, Pennsylvania 16802, USA}
\affiliation{School of Physics and Astronomy, Cardiff University, Cardiff, CF24 3AA, United Kingdom}


\date{\today}

\begin{abstract}
Next-generation terrestrial gravitational-wave observatories will detect $\mathcal{O}(10^{5})$ signals from compact binary coalescences every year. These signals can last for several hours in the detectors' sensitivity band and they will be affected by multiple unresolved sources contributing to a confusion-noise background in the data. Using an information-matrix formalism, we estimate the impact of the confusion noise power spectral density in broadening the parameter estimates of a GW170817-like event.  If our estimate of the confusion noise power spectral density is neglected, we find that masses, spins, and distance are biased in about half of our simulations under realistic circumstances.  The sky localization, while still precise, can be biased in up to $80\%$ of our simulations, potentially posing a problem in follow-up searches for electromagnetic counterparts.
\end{abstract}

\maketitle

\noindent {\bf \em Introduction.} 
The LIGO-Virgo-Kagra (LVK) observatories~\cite{Harry:2010zz, LIGOScientific:2014pky,VIRGO:2014yos,Aso:2013eba}
have opened up the era of gravitational-wave (GW) astronomy, detecting up to $100$ compact binary coalescence events~\cite{LIGOScientific:2021djp,LIGOScientific:2021psn,Nitz:2021zwj,Olsen:2022pin}. Next-generation (XG) interferometers such as Cosmic Explorer (CE)~\cite{Reitze:2019iox} and the Einstein Telescope (ET)~\cite{Punturo:2010zz} are expected to detect $\mathcal{O}(10^{5})$ events per year with unprecedented signal-to-noise ratios (SNRs)~\cite{Borhanian:2022czq,Pieroni:2022bbh,Iacovelli:2022bbs}. The improved sensitivity at lower frequencies implies that GW signals will last longer in the detector sensitivity band, allowing for timely source localization for potential electromagnetic follow-up campaigns~\cite{Ronchini:2022gwk}.

Longer durations, however, imply the presence of multiple unresolved signals that overlap with a given detected source signal, producing a confusion background~\cite{Antonelli:2021vwg,Pizzati:2021apa,Samajdar:2021egv}.
The possibility of detecting confusion noise from unresolved compact binary signals in the context of XG detectors was first considered in Ref.~\cite{Regimbau:2009rk}. More recent work studied how this confusion noise would limit the redshift reach of XG detectors~\cite{Wu:2022pyg}. Here we focus on the impact of confusion noise from multiple overlapping sources on the {\em parameter estimation} of resolved sources in XG detectors.

Long resolved signals will overlap with a large number of background sources and will be most affected by confusion noise. At leading order, the duration of a signal in the detector sensitivity band is given by~\cite{Peters:1963ux,Sathyaprakash:2009xs,Pizzati:2021apa}
\beq
T_{\rm det}
= 25851\left(\frac{\mathcal{M}_{\rm z}}{1.185\,\rm{M}_\odot}\right)^{-5/3}\left(\frac{f_0}{3\,\rm{Hz}}\right)^{-8/3}\,\rm{s}\,,
  \label{eq:duration}
\eeq
where $\mathcal{M}_{\rm z}=\mathcal{M}(1+z)$ is the detector-frame chirp mass, $z$ is the redshift of the source, $f_0$ is the starting frequency of the signal in the detector sensitivity band, and we have chosen the reference value of $\mathcal{M}_{\rm z}$ to match a {\em golden} GW170817-like  binary neutron star (BNS)~\cite{LIGOScientific:2017vwq}.
Thus, signals generated by low-mass, low-redshift sources, such as close-by BNSs, are most affected by confusion noise~\cite{Pizzati:2021apa,Samajdar:2021egv}.
The signal duration is very sensitive to the starting frequency: a GW170817-like binary would be in band for about
$\sim 45\,\rm{min}$,
$2\,\rm{hours}$,
and
$7\,\rm{hours}$
for $f_0=7\,\rm{Hz}$, $5\,\rm{Hz}$, and $3\,\rm{Hz}$, respectively. 

The unresolved background is expected to be dominated by BNSs~\cite{LIGOScientific:2021psn,Renzini:2022alw,Zhou:2022otw,Zhou:2022nmt}, which tend to have longer durations than binaries involving black holes, and hence a higher probability of overlap~\cite{Pizzati:2021apa,Samajdar:2021egv}. 
Depending on the local BNS merger rate, which is currently uncertain~\cite{LIGOScientific:2021psn}, the rate of BNS coalescences in the Universe can range from 1 every $\sim 30\,\rm{min}$ to 1 every $\sim 10\,\rm{s}$~\cite{Pizzati:2021apa,Borhanian:2020ypi}. Assuming a fraction of detected events between $30\%$ and $60\%$ (depending on factors such as population models, detection thresholds, etcetera), it is clear that dealing with several simultaneous BNS signals, both resolved and unresolved, is crucial for data analysis purposes in XG detectors.

In this work, we use metrics devised in Ref.~\cite{Antonelli:2021vwg} within an information-matrix formalism
to estimate the impact of confusion noise on the parameter estimation of a loud BNS signal with XG detectors. 
We generate a background of unresolved BNS signals from state-of-the-art population models~\cite{LIGOScientific:2021djp,LIGOScientific:2021psn},  and we compute a ready-to-use estimate of the confusion noise power spectral density (PSD) for parameter estimation in XG detectors. 

Throughout this work, we use the public package \textsc{gwbench}~\cite{Borhanian:2020ypi} to compute waveforms, SNRs, and information matrices.
We adopt the $\Lambda$CDM cosmological model with parameters taken from Planck 2018~\cite{Planck:2018vyg}.

\noindent {\bf \em Astrophysical population and setup.} 
We consider a fiducial XG 3-detector network consisting of a $40$-km scale CE located in Idaho (USA), a $20$-km CE in Australia, and an ET in Italy~\cite{Borhanian:2020ypi}. We include Earth-rotation effects in the antenna patterns for all the signals in our study.

In order to generate the BNS background, we employ a population model consistent with the latest LVK GWTC-3 catalog~\cite{LIGOScientific:2021psn,LIGOScientific:2020kqk}. The source-frame component masses $m_1$ and $m_2$ of each BNS are sampled using the preferred model from Ref.~\cite{Farrow:2019xnc}, where the primary mass follows a double Gaussian distribution and the secondary mass is sampled uniformly.
We adopt the same BNS redshift distributions as in Ref.~\cite{Borhanian:2022czq}. We assume that the binary formation rate follows the Madau-Dickinson~\cite{Madau:2014bja} cosmic star formation rate (SFR), and we obtain the merger rate by convolving the SFR with a standard $p(t_d)\propto 1/t_d$ time-delay distribution~\cite{Dominik:2013tma, LIGOScientific:2016fpe, LIGOScientific:2017zlf,Meacher:2015iua, LIGOScientific:2017zlf}.
Estimates of the BNS background are affected by large astrophysical uncertainties. We set the overall normalization of the merger rate by adopting a fiducial local merger rate of $320\,\rm Gpc^{-3}yr^{-1}$, corresponding to the best estimate from the GWTC-2 catalog~\cite{LIGOScientific:2020kqk}. This rate is still consistent with the local rate estimates from the GWTC-3 catalog~\cite{LIGOScientific:2021psn,LIGOScientific:2020kqk}, while also allowing for a more immediate comparison with recent forecasts for XG observatories~\cite{Borhanian:2022czq,Ronchini:2022gwk}. 
We estimate the impact of the local merger rate uncertainty by varying it within the $90\%$ confidence interval $[10,1700]\,\rm Gpc^{-3}yr^{-1}$ inferred from the GWTC-3 catalog~\cite{LIGOScientific:2021psn}.

For all the GW signals in the background,  we assume both the sky positions of the sources and their orientation with respect to the observer's line of sight to be isotropic.
The polarization angle $\psi$ is drawn uniformly in the range $[0, 2\pi]$.
We assume nonspinning binaries and neglect tidal deformabilities, as these parameters have subdominant effects on background estimates.

We assess the detectability of GW signals by computing their optimal SNR $\rho = (h|h)$, where $(\cdot|\cdot)$ is the usual signal inner product~\cite{Maggiore:2007ulw,Schutz:2011tw}
\beq
(a|b) = 4\Re\int^{\infty}_{0}\frac{\tilde{a}(f)\tilde{b}^*(f)}{S_{\rm n}(f)}\,df \,.
\label{eq:innerprod}
\eeq
Here we denote signals in the frequency domain with a tilde, complex conjugates with an asterisk, and $S_{\rm n}(f)$ is the instrumental-noise PSD. We assume the noise to be uncorrelated between different detectors, hence the network SNR is given by summing the single-detector SNRs in quadrature~\cite{Cutler:1994ys}. We consider a signal to be resolved if its network SNR is above a threshold $\rho_{\rm thr}=12$. Any signal in our catalog with SNR below this threshold contributes to the background.

We perform parameter inference on a golden BNS similar to GW170817~\cite{LIGOScientific:2017vwq}, with detector-frame chirp mass $\mathcal{M}_z=1.195508~M_{\odot}$, symmetric mass ratio $\eta=0.2489$, (aligned) dimensionless spins $ \chi_{1}=\chi_{2}=0$, and tidal deformability parameters $\tilde\Lambda=600$ and $\delta\tilde\Lambda=0$ (see~\cite{Wade:2014vqa}).
We place the source at a luminosity distance of $D_{\rm L}= 40.152~{\rm Mpc}$, with sky location, polarization, and inclination angles $\alpha=\delta=\psi=\iota=\pi/4$. We fix the time and phase at coalescence to $t_{\rm c}=\phi_{\rm c}=0$. Below we will denote this set of ``true'' injected parameters by $\vec{\theta}_{\rm tr}$.

We simulate all the signals up to a frequency of $2048\,\rm{Hz}$. The detector noise makes signal contributions above this maximum frequency negligible~\cite{Borhanian:2022czq}. 
The starting frequency $f_0$ plays a crucial role as it dictates the duration of the signals in the detectors' sensitivity bands, and thus determines whether the source will be affected by low-frequency, subthreshold overlapping signals. We assess the impact of this parameter by considering three different values in our study: $f_0=3\,\rm{Hz}$, $5\,\rm{Hz}$, and $7\,\rm{Hz}$.

For the golden BNS signal we use the waveform model \texttt{IMRPhenomD\_NRTidalv2}~\cite{Dietrich:2019kaq}, which includes tidal effects. With this choice, the SNR of this signal in our detector network and in the absence of confusion noise is $\rho_{\rm det}=1679$, $1674$, and $1634$ for starting frequencies $f_0=3\,\rm{Hz}$, $5\,\rm{Hz}$, and $7\,\rm{Hz}$, respectively. For the background signals we use the inspiral-only waveform model \texttt{TaylorF2}~\cite{Sathyaprakash:1991mt,Buonanno:2009zt},
since we expect most BNS signals in XG detectors to be inspiral dominated.

To determine which background signals overlap with the golden BNS, we compute the inspiral time-frequency evolution of each of them at $3.5$ post-Newtonian order using the public package \texttt{PyCBC}~\cite{Usman:2015kfa}.
We assign a fixed time of arrival to the reference BNS, generate a catalog of BNS signals up to redshifts $z=10$ from our population model, and assign an arrival time to each of the undetected signals. We conservatively draw the arrival times uniformly over the span of a week around the arrival time of the golden binary. After estimating its duration, each background signal has a definite starting time $t_0$ and ending time $t_1$ in the network sensitivity band. If the time interval $[t_0,t_1]$ associated to a background signal overlaps with the time interval $[t_0^{\rm det}, t_1^{\rm det}]$ associated with the detected BNS, then the background signal contributes to the confusion noise.
Signals can overlap completely or partially. We use our time-frequency evolution to select, for each overlapping signal, the portion in the frequency domain that actually overlaps with the golden BNS. 

The superposition of the unresolved frequency chunks that overlap with the detected BNS signal generates a confusion noise
\beq
\Delta\tilde{H}(f,\vec{\theta}^i,\vec{\theta}_{\rm tr}) = \sum_{i=1}^{N_{\rm over}}\tilde{h}_{\rm unres}^i(f;\vec{\theta}^i) \,,
\label{eq:confnoise}
\eeq
where $\tilde{h}_{\rm unres}^i(f;\vec{\theta}^i)$ are the $N_{\rm over}$ unresolved chunks of background signals that overlap with the detected signal, and $\vec{\theta}^i$ their parameters. By averaging over multiple realizations, we find a median number of overlapping unresolved signals of $N_{\rm over}=302$, $77$, and $32$ for starting frequencies of $f=3\,\rm{Hz}$, $5\,\rm{Hz}$, and $7\,\rm{Hz}$, respectively.

\noindent {\bf \em Confusion noise PSD.} 
\begin{figure*}[t]
\includegraphics[width=\textwidth]{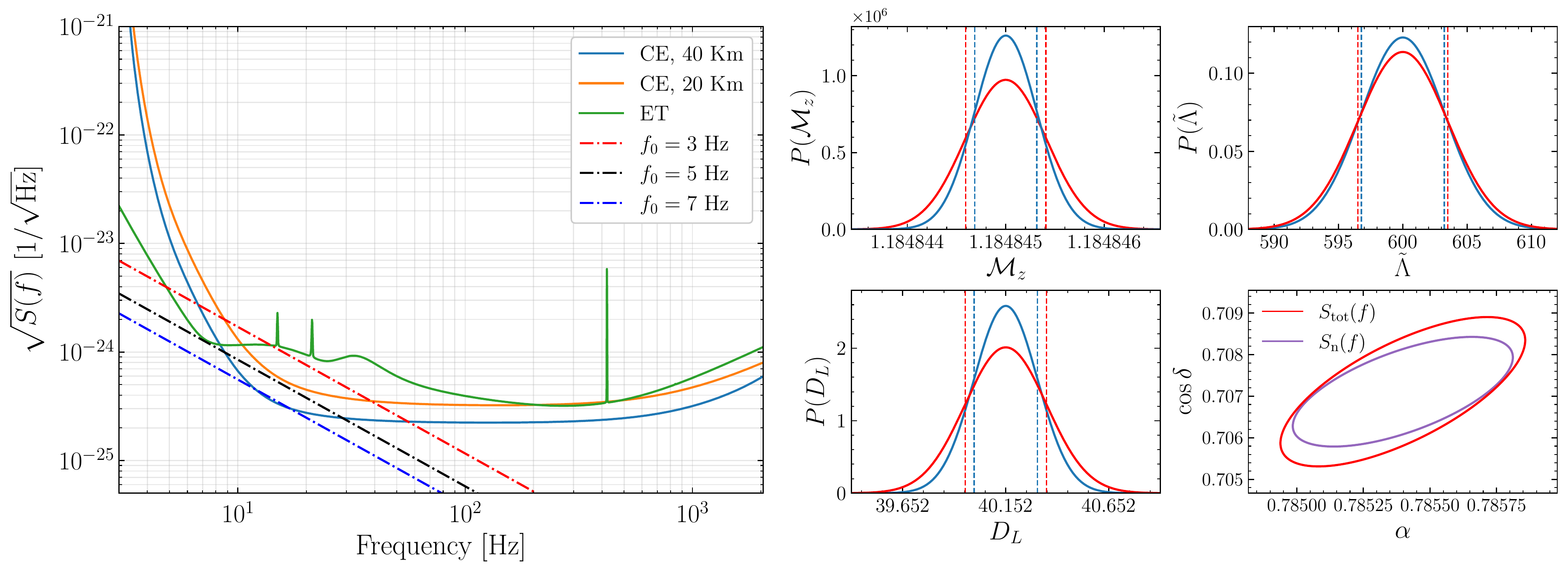}
\caption{Left panel: comparison between the amplitude spectral density (ASD) $\sqrt{S_{\rm n}(f)}$ associated to the detector noise for each interferometer, and that resulting from the confusion noise $\sqrt{S_{\rm conf}(f)}$ for different starting frequencies. For visualization purposes, we show only the confusion noise ASDs for the $40-$km CE. Results for the other detectors are similar. Right panel: comparison between the uncertainties on different binary parameters computed with the detector noise PSD $S_n(f)$ (blue line) and the total PSD $S_{\rm tot}(f)$ (red line). The vertical lines in the $1$D posteriors and the contour in the bottom-right panel represent $68\%$ credible levels. The curves are computed for $f_0=3\,\rm{Hz}$ and for the fiducial local merger rate of $320\,\rm Gpc^{-3}yr^{-1}$; in this case, the relative SNR loss when we include the confusion noise PSD is 30\%. 
}
\label{fig:psd}
\end{figure*}
The presence of confusion noise $\Delta \tilde{H}$ acts as an additive contribution to the GW detector output
\beq
\tilde{d}(f) = \tilde{h}_{\rm det}(f;\vec{\theta}_{\rm tr}) + \tilde{n}(f)+\Delta \tilde{H}(f) ,
\eeq
where $\tilde{h}_{\rm det}(f;\vec{\theta}_{\rm tr})$ is a single detected signal with true parameters $\vec{\theta}_{\rm tr}$ and $\tilde{n}(f)$ is the usual instrumental noise, modelled by a PSD $S_{\rm n}(f)$ via the equation~\cite{Maggiore:2007ulw}
\beq
\langle \tilde{n}(f)\tilde{n}^*(f') \rangle = \frac{1}{2}\delta(f-f')S_{\rm n}(f) \,.
\label{eq:detpsd}
\eeq
Here, $\delta$ is the Dirac delta, and $\langle\cdot\rangle$ is an ensemble average over multiple noise realizations. 
If the confusion noise is Gaussian and stationary, it can be fully characterized by a PSD $S_{\rm conf}(f)$ defined in an analogous way~\cite{Antonelli:2021vwg}, i.e.,
\beq
\langle \Delta\tilde{H}(f)\Delta\tilde{H}^*(f') \rangle = \frac{1}{2}\delta(f-f')S_{\rm conf}(f) \,.
\label{eq:psdconf}
\eeq
We produce $2400$ realizations of confusion noise as described previously, each of them generated  from a different random sample from our population model.

For a confusion background of GW signals from inspiraling compact binary coalescences (as in the case of our study), the confusion noise PSD is well approximated by a power law~\cite{Maggiore:1999vm,Regimbau:2016ike}
\beq
S_{\rm conf}(f)=A_{\rm ref}\,\left(\frac{f}{f_{\rm ref}}\right)^{-7/3}\,,
\label{eq:powerlawpsd}
\eeq
where $f_{\rm ref}$ is an arbitrary reference frequency. We fit this power law with the confusion noise realizations produced above, up to a frequency cutoff of $50\,\rm{Hz}$. We extrapolate our power-law fit to higher frequencies, where we expect the instrumental noise to dominate over the confusion noise~\cite{Wu:2022pyg}.

The results are shown in the left panel of Fig.~\ref{fig:psd}, where we compare the fitted confusion-noise PSDs $S_{\rm conf}(f)$ for different starting frequencies with the detector-noise PSDs. For visualization purposes, we only show the confusion noise PSD for the $40$-km CE, but results for the other detectors are similar, since we adopt the same starting frequencies for all of them. From Eqs.~\eqref{eq:duration} and \eqref{eq:psdconf}, the amplitude of the confusion noise PSD must satisfy the scaling relations
$A_{\rm ref} \propto N_{\rm over} \propto T_{\rm det} \propto f_0^{-8/3}$.
At a reference frequency of $f_{\rm ref}=20\,\rm{Hz}$ we find the amplitudes
\begin{equation}
    A_{\rm ref}=
    \begin{cases}
      5.8\times10^{-49}, & \text{for } f_0=3\,\rm{Hz} \\
      1.4\times10^{-49}, & \text{for } f_0=5\,\rm{Hz} \\
      6.2\times10^{-50}, & \text{for } f_0=7\,\rm{Hz}
    \end{cases},
\end{equation}
in excellent agreement with this scaling. Depending on the starting frequency, the confusion noise PSD becomes comparable to (or even dominant over) the instrumental noise at low frequencies. Therefore, uncertainties on the inferred source parameters will increase in comparison to neglecting the effects of the confusion noise.

We compute the uncertainties on the $13$ BNS parameters using the information matrix $\Gamma_{ij}=(\del_i h|\del_j h)$~\cite{Finn:1992wt,Schutz:2011tw}, where $\del_i$ denotes the partial derivative with respect to the $i$-th parameter. Within this formalism, the uncertainty on the inferred parameters can be estimated as $\sigma^i=\sqrt{(\Gamma^{-1})^{ii}}$ adopting a given PSD when computing the inner product of Eq.~\eqref{eq:innerprod}.

\begin{figure*}[t]
\includegraphics[width=\textwidth]{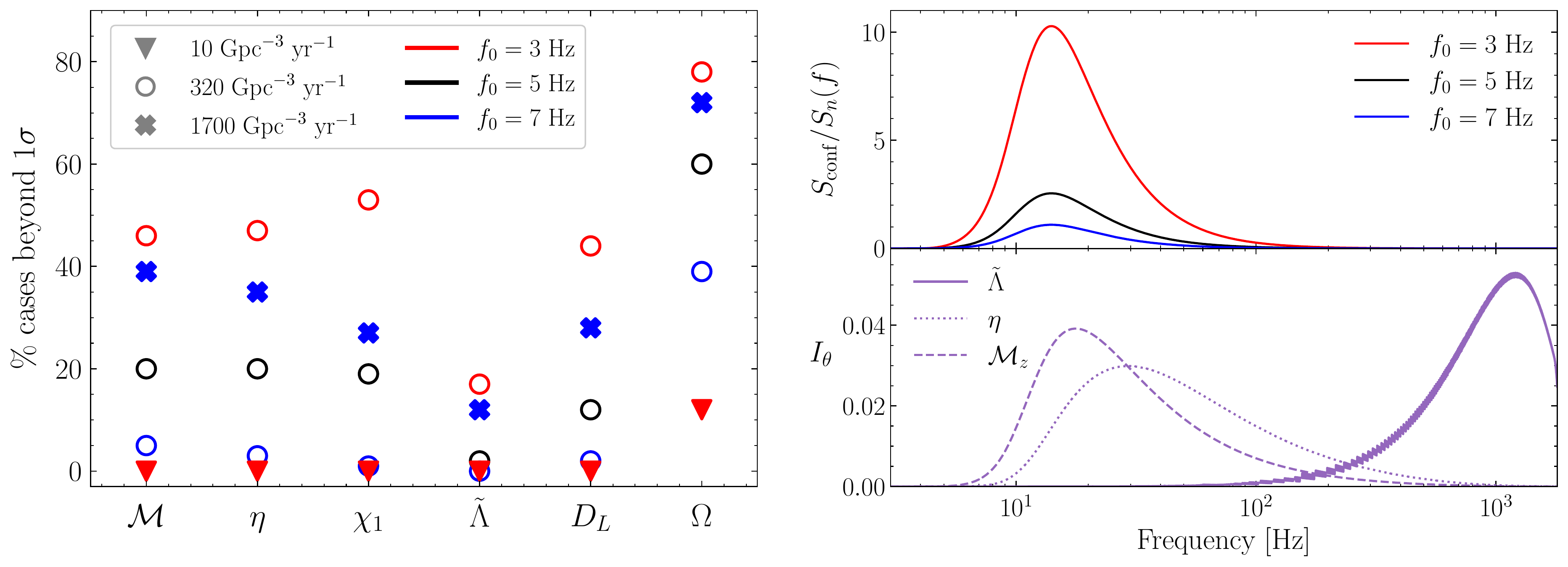}
\caption{Left panel: percentage of the realizations of the confusion noise for which the parameters $(\mathcal{M}, \eta, \chi_1, \tilde{\Lambda}, D_\mathrm{L}, \Omega)$ are biased at $68\%$ level.  Different markers and colors represent different starting frequencies and local merger rates, respectively. Top-right panel: ratio between confusion noise and instrumental noise PSDs for different starting frequencies. Bottom-right panel: measurability integrands as functions of frequency for $\mathcal{M}_z$ ,$\eta$ and $\tilde{\Lambda}$. For visualization purposes, we show only the PSD ratios and measurability integrands for the $40-$km CE. The overall scalings with frequencies for other detectors are similar.}
\label{fig:errors}
\end{figure*}
Assuming no correlation between instrumental and confusion noise, the total noise PSD is given by $S_{\rm tot}(f)\equiv S_{\rm n}(f)+S_{\rm conf}(f)$.
We can compute the total uncertainties by plugging $S_{\rm tot}(f)$ into Eq.~\eqref{eq:innerprod} for the information matrix, and the uncertainties due to instrumental noise alone by considering just $S_{\rm n}(f)$. The right panels of Fig.~\ref{fig:errors} show the comparison between Gaussian error distributions with standard deviations estimated from information matrices using either  $S_{\rm n}(f)$ only (in blue) or the total PSD $S_{\rm tot}(f)$ (in red). The distributions shown are for $f_0=3\,\rm Hz$ and for the fiducial local merger rate of $320\,\rm Gpc^{-3}yr^{-1}$. In general, the contribution of confusion noise is significant, increasing the errors on all parameters between $10\%$ and $40\%$. For crucial parameters such as luminosity distance and sky position the $68\%$ uncertainties increase by factors of $\sim30\%$ and $\sim40\%$, respectively, with potential impact on electromagnetic follow-up campaigns or in the estimation of cosmological parameters. 

\noindent {\bf \em Confusion noise errors.} 
If ignored, the confusion noise term causes biases when inferring the parameters of the resolved source.
Reference~\cite{Antonelli:2021vwg} proposed a formalism to calculate the additional shift in the inferred parameters caused by a certain confusion noise realization within the linear signal approximation:
\beq
\Delta\theta^i_{\rm conf} = (\Gamma_{\rm n}^{-1})^{ij}\left(\del_j h\middle| \Delta H \right) \,,
\label{eq:deltatheta_conf}
\eeq
where $\Gamma_{\rm n}$ is the information matrix associated to the instrumental noise PSD $S_{\rm n}(f)$, and we are summing over repeated indices. A metric to assess whether the confusion noise dominates over  the detector noise for parameter estimation can further be defined by the $1\sigma$-ratio
$\mathcal{R}^i = \Delta\theta^i_{\rm conf}/\sigma^i_{\rm n}$,
where $\sigma_{\rm n}^i=\sqrt{(\Gamma_{\rm n}^{-1})^{ii}}$ is the typical statistical error due to instrumental noise~\cite{Antonelli:2021vwg}. If $|\mathcal{R}^i| \lesssim 1$, the shift induced by $\Delta\theta^i_{\rm conf}$ is consistent with the $1\sigma$ statistical deviation, while the confusion background becomes dominant for parameter estimation for $|\mathcal{R}^i| > 1$. 

It is important to note that both $\Delta\theta^i_{\rm conf}$ and $\sigma^i_{\rm stat}$ scale in the same way with the SNR of the detected signal (i.e., as $\sim 1/\rho_{\rm det}$). As a consequence,
$\mathcal{R}^i$ does not depend on $\rho_{\rm det}$. What is crucial to understand whether the confusion noise dominates over the statistical uncertainty is instead the SNR of the confusion background, $\rho_{\rm conf}^2 = (\Delta H | \Delta H) $, which provides an estimate of how ``loud'' $\Delta \tilde{H}$ is compared to the detector noise $n$: if $\rho_{\rm conf}$ is large, the noise floor is dictated by the confusion background, and thus the confusion noise error is larger than the typical statistical error. The SNR of the confusion background is ultimately determined by how many unresolved signals overlap with the detected signal $h_{\rm det}$, and for how long. This, in turn, depends on the duration of $h_{\rm det}$ in band. 

The left panel of Fig.~\ref{fig:errors} shows the percentage of cases with $|\mathcal{R}^i| > 1$
for several binary parameters. These percentages are shown for different values of the starting frequency and of the local merger rate. We find that the tidal deformability $\tilde\Lambda$ is less affected than other parameters, while the sky position $\Omega$ is the most affected. Moreover, the number of biased cases is highly dependent on the local merger rate: for a low local rate of $10\,\rm Gpc^{-3}yr^{-1}$ we find no biased cases even when $f_0=3\,\rm{Hz}$, while for a rate of $1700\,\rm Gpc^{-3}yr^{-1}$ there is a significant fraction of biased cases even when $f_0=7\,\rm{Hz}$.

The different impact of confusion noise on the inference of the various binary parameters is clarified by the right panels of Fig.~\ref{fig:errors}. In the top-right panel we show the ratio between the confusion noise PSD $S_{\rm conf}(f)$ and the instrumental noise PSD $S_{\rm n}(f)$ for our fiducial local merger rate and different starting frequencies. This ratio is mostly different from zero at low frequency, peaking at about $15\,\rm{Hz}$, in agreement with recent work~\cite{Wu:2022pyg}. In the bottom-right panel we plot the (normalized) integrand of selected diagonal elements $\Gamma^{ii}$ of Eq.~\eqref{eq:innerprod}. This quantity , also known as the ``measurability integrand'' $I_{\theta}$~\cite{Damour:2012yf}, allows us to understand which frequency regions contribute the most to the measurability of parameter $\theta$. The normalized measurability integrands for $\mathcal{M}_z$, $\eta$, and $\tilde{\Lambda}$ were computed using the detector noise PSD $S_{\rm n}(f)$. By direct comparison with the PSD ratio, we can estimate which parameter errors are most affected by confusion noise. Parameters whose integrands are large in a frequency range where $S_{\rm conf}(f)$ dominates over $S_{\rm n}(f)$, such as $\mathcal{M}_{\rm z}$ or $\eta$, are most affected. Parameters that are mostly measured at high frequencies, such as $\tilde\Lambda$, are less affected.

\noindent {\bf \em Conclusions and future directions.} 
The main conclusion to be drawn from this work is that confusion noise in XG detectors plays an important role when inferring the parameters of low-redshift, low-mass sources, such as close-by BNSs. We have estimated the confusion noise PSD from a BNS background for a golden BNS with the same masses and redshift as GW170817. Neglecting confusion noise can lead to significant biases in the inferred parameters, especially in the sky position. The impact of confusion noise depends crucially on (i) the low-frequency sensitivity limit of the detectors, and (ii) the large uncertainty in the local merger rate of BNSs.

Even when including confusion noise, the parameter estimation uncertainties of a GW170817-like BNS in a network of XG observatories are still orders of magnitude better than current interferometers: for example, the $68\%$ uncertainty in the sky area goes from $\sim4\times 10^{-3}\,\rm{deg}^2$ to $\sim6\times 10^{-3}\,\rm{deg}^2$ when we include confusion noise. More work is needed to understand how confusion noise in XG detectors will impact our ability to provide alerts for electromagnetic counterparts.
Our confusion noise estimates are somewhat conservative, as we do not include subdominant backgrounds from binary black holes or neutron star-black hole binaries, and we assume perfect subtraction of other detected signals.
It will be important to explore data-analysis techniques to reduce the impact of confusion noise on parameter estimation for XG detectors. Some possibilities include global-fit schemes~\cite{Littenberg:2020bxy}, notching in time-frequency space~\cite{Zhong:2022ylh}, and Bayesian techniques to estimate the foreground and background signal parameters simultaneously~\cite{Biscoveanu:2020gds}.

\noindent {\bf \em Acknowledgements.} 
We thank Hsin-Yu Chen, Konstantinos Kritos and Ken ``Enzo'' Ng for insightful discussions. L.R., A.A., R.C., M.\c{C}. and E.B. are supported by NSF Grants No. AST-2006538, PHY-2207502, PHY-090003 and PHY20043, and NASA Grants No. 19-ATP19-0051, 20-LPS20-0011 and 21-ATP21-0010. M.\c{C}.\ is also supported by Johns Hopkins University through the Rowland Research Fellowship.
S.B. acknowledges support from the Deutsche Forschungsgemeinschaft, DFG, project MEMI number BE 6301/2-1.
B.S.S. is supported by NSF Grants No. AST-2006384, PHY-2012083 and PHY-2207638. Part of E.B.’s and B.S.S.’s work was performed at the Aspen Center for Physics, which is supported by National Science Foundation grant PHY-1607611. This research was also supported in part by the National Science Foundation under Grant No.  NSF PHY-1748958.
This research project was conducted using computational resources at the Maryland Advanced Research Computing Center (MARCC).
The authors also acknowledge the Texas Advanced Computing Center (TACC) at The University of Texas at Austin for providing HPC resources that have contributed to the research results reported within this paper. URL: \url{http://www.tacc.utexas.edu}~\citep{10.1145/3311790.3396656}.

\bibliography{confusion_noise_prl}

\end{document}